\documentclass[manuscript]{aastex}

\RequirePackage[mathscr]{eucal}
\usepackage[T1]{fontenc}
\usepackage{lmodern}

\newcommand{\zitamail}{zita@evergreen.edu}

\newcommand{\Bbar}{\overline{\bf{B}}}

\newcommand{\curle}{\mathscr{E}}

\newcommand{\vbar}{\overline{\bf{v}}}

\newcommand{\half}{\frac{1}{2}}
\newcommand{\BB}{\bf{B}}
\newcommand{\bp}{\bf{b'}}
\newcommand{\vp}{\bf{v'}}
\newcommand{\gradeta}{\nabla \eta}
\newcommand{\Rud}{R\"{u}diger}
\newcommand{\sqe}{\sqrt{\eta}}
\newcommand{\sqet}{\sqrt{\eta_T}}

\shorttitle{Magnetic Advection by Diffusivity Gradients}
\shortauthors{Zita}

\begin{document}

\title{Analytic Origin of Advection of Magnetic Fields by
Diffusivity Gradients}


\author{E.J. Zita}
\affil{The Evergreen State College, Olympia, WA, 98507}
\email{\zitamail}

%

\begin{abstract}
We derive the advection of magnetic fields due to
gradients in magnetic diffusivity, starting from
the magnetic induction equation.
We discuss physical examples,
and compare our results to those in the literature.
Our induction and diffusion equations may be useful in
solar dynamo studies, and for space and laboratory plasmas.
\end{abstract}


\section{Introduction}
Dynamic magnetic fields have complex and interesting relationships with
a physical property of plasmas, the magnetic diffusivity, $\eta$,
defined as the ratio of the electrical resistivity
to the magnetic permeability.
The usual form of the induction equation
derived from Maxwell's equations and Ohm's law:
\begin{equation}
\label{eqn1}
\frac{d\bf{B}}{dt}=\nabla \times \left[ (\bf{v} \times \bf{B}) -
\eta \nabla \times \bf{B} \right]
\end{equation}
shows that both diffusivity $\eta$ and flows $\bf{v}$ constrain the
rate of change of magnetic fields $\BB$.

Magnetic diffusivity can be enhanced by plasma turbulence (Leighton 1964),
and strong magnetic fields can suppress turbulence
\citep{RK, tob06}.
Therefore magnetic fields can suppress (or ``quench'')
turbulence-enhanced diffusivity (Gilman \& Rempel 2005),
which can then permit fields to amplify locally, within limits.
Nonlinear interactions between diffusivity quenching and magnetic fields evolution 
are being studied 
numerically in the context of the solar dynamo \citep{GG09}.

In this paper we are primarily
interested in an analytic relationship between dynamic magnetic fields $\BB$ and $\eta$.
We find that gradients in magnetic diffusivity $\eta$ can cause the
magnetic fields $\BB$ to advect, adding a new flow term to the induction equation.
Our new advection term can ultimately restructure both $\eta$
(which generally varies in space as well as in time) and the magnetic fields.
Such processes may have important physical implications for magnetohydrodynamics
in the convection zone and for the solar dynamo, among other systems.

Our original motivation in this inquiry was to better understand the fundamental source
of the ``diamagnetic pumping velocity,'' $U^{dia} =f( \nabla \eta_T)$ described 
as driving magnetic advection in the context of  
the mean field dynamo equation
\begin{equation}
\label{rudMF}
\frac{d\Bbar}{dt}=\nabla \times \left[ \vbar \times \Bbar
+ \alpha \cdot \Bbar
-\sqet \, \nabla \times \left( \sqet \Bbar \right) \right]
\end{equation}
where $\eta_T$ is the linear turbulent magnetic diffusivity \citep{RH}, Ch.4.
In the process of investigating this relation, 
we rigorously derived expressions for magnetic field advection, as described below.
\section{Derivation of magnetic advection from diffusivity gradients in the induction equation}
\label{bozomath}
We derive magnetic advection explictly from diffusivity gradients, using the
induction equation (1).
In an effort to understand \Rud's equation (\ref{rudMF}),
we note the presence of $\sqe$ terms and seek to relate them directly
to $\gradeta$.
Since
$\frac{d}{dr}\sqrt{\eta} = \half \frac{1}{\sqrt{\eta}} \frac{d}{dr} \eta \, ,$
%
we can rewrite this as
\begin{equation}
\label{2sqrteta}
\nabla \sqrt{\eta} = \frac{\nabla \eta}{2 \sqrt{\eta}} ,
\end{equation}
assuming that diffusivity gradients are only in the radial direction.
It will turn out to be useful to expand a term similar to the diffusivity term
in the induction equation (1),
using the result above and standard vector identities.
Letting $\bf{C} = \nabla \times \bf{B}$, we find that
\begin{eqnarray*}
\nabla \times \left[\sqrt{\eta} \, \nabla \times \bf{B} \right]
&=& \nabla \times \left[ \sqrt{\eta} \,\bf{C} \right] \\
&=& \sqrt{\eta} \nabla \times \bf{C} + \nabla \sqrt{\eta} \times \bf{C} \\
&=& \sqrt{\eta}\, \nabla \times (\nabla \times \bf{B} ) +
\nabla \sqrt{\eta} \times (\nabla \times \bf{B}) .\\
\end{eqnarray*}
Using (\ref{2sqrteta}) in the last term, this becomes
\begin{equation}
\label{3sqrtetaNVI}
\nabla \times \left[\sqrt{\eta} \, \nabla \times \bf{B} \right]
= \sqrt{\eta}\, \nabla \times (\nabla \times \bf{B} ) +
\frac{\nabla \eta}{2 \sqrt{\eta}} \times (\nabla \times \bf{B}).
\end{equation}
Recalling the identity $\nabla \times \nabla \times \BB = - \nabla^2 \BB$,
and multiplying by $\sqrt{\eta}$, we find
\begin{equation}
\label{4del2B}
\sqrt{\eta}\left( \nabla \times \left[\sqrt{\eta}\, \nabla \times \bf{B} \right] \right)
= - \eta \nabla^2 \BB +
\half \nabla \eta \times (\nabla \times \bf{B}) .
\end{equation}
The last term above can be rewritten using the vector identity
\begin{equation}
\bf{D} \times (\nabla \times \bf{B}) =
 \nabla(\bf{D} \cdot \bf{B}) - \bf{B} \times (\nabla \times \bf{D}) -
 (\bf{D} \cdot \nabla) \bf{B} - (\bf{B} \cdot \nabla) \bf{D}
\end{equation}
and letting $\bf{D} = \nabla \eta$:
\begin{equation}
\nabla \eta \times (\nabla \times \bf{B}) =
 \nabla(\nabla \eta \cdot \bf{B}) - \bf{B} \times (\nabla \times \nabla \eta) -
 (\nabla \eta \cdot \nabla) \bf{B} - (\bf{B} \cdot \nabla) \bf{\gradeta}.
\end{equation}
Keeping only first-order differentials in $\eta$, we have
\begin{equation}
\label{5curlcurl}
\nabla \eta \times (\nabla \times \bf{B}) =
 - (\gradeta \cdot \nabla) \BB \, .
\end{equation}
Combining this result with (\ref{4del2B}) yields
\begin{equation}
\label{6del2B}
\sqrt{\eta}\left( \nabla \times \left[\sqrt{\eta}\, \nabla \times \bf{B} \right] \right)
= -\eta \nabla^2 \BB - \half (\nabla \eta \cdot \nabla) \BB \, .
\end{equation}
Similarly, we can rewrite the exact diffusivity term in the induction equation (1),
using
$\bf{C} = \nabla \times \bf{B}$, the same vector identities,
and (8):
\begin{eqnarray*}
\nabla \times \left[ \eta \nabla \times \bf{B} \right]
&=& \eta \nabla \times (\nabla \times \bf{B})  +
\nabla \eta \times (\nabla \times \bf{B}) \\
&=& -\eta \nabla^2 \BB - (\nabla \eta \cdot \nabla) \BB . \\
\end{eqnarray*}
Halving the last term in this result, we obtain two terms as in the right-hand side of (\ref{6del2B}):
\begin{equation}
\label{7halfhalf}
\nabla \times \left[ \eta \nabla \times \bf{B} \right]
= - \eta \nabla^2 \BB - \half (\nabla \eta \cdot \nabla) \BB
- \half (\nabla \eta \cdot \nabla) \BB .
\end{equation}
Comparing (\ref{6del2B}) and (\ref{7halfhalf}), we find
\begin{equation}
\label{8combine}
\nabla \times \left[ \eta \nabla \times \bf{B} \right]
= \sqrt{\eta}\left[ \nabla \times \left(\sqrt{\eta}\, \nabla \times \bf{B} \right) \right]
- \half (\nabla \eta \cdot \nabla) \BB  \, .
\end{equation}
Using (3), we can rewrite
\begin{displaymath}
\half (\nabla \eta \cdot \nabla) = \nabla \sqrt{\eta} \cdot (\sqrt{\eta}\,\nabla) \, .
\end{displaymath}
Then (\ref{8combine}) becomes
\begin{equation}
\label{9almostDE}
\nabla \times \left[ \eta \nabla \times \bf{B} \right]
=\sqrt{\eta}\left[ \nabla \times \left(\sqrt{\eta} \,
\nabla \times \bf{B} \right) \right]
- \nabla \sqrt{\eta} \cdot (\sqrt{\eta}\,\nabla) \BB \, .
\end{equation}
A more enlightening form can be obtained by letting
$\nabla' = \sqrt{\eta} \, \nabla \, ,$ and
$\bf{U_{\eta}} \equiv \nabla \sqrt{\eta}$ \.:
\begin{equation}
\label{10de}
\nabla \times \left[ \eta \nabla \times \bf{B} \right]
= \nabla' \times
\nabla' \times \bf{B}
- ( \bf{U_{\eta}} \cdot  \nabla' \BB) \, ,
\end{equation}
where $\bf{U_{\eta}}$ is the magnetic field advection velocity
due to diffusivity gradients, and
the gradient operator $\nabla'$ is also rescaled by the square root of the
diffusivity.
Note that (\ref{10de}) has the form of a
diffusion equation itself, with constant diffusion coefficient equal to 1,
which may be particularly easy to simulate.
Finally, the induction equation becomes
\begin{eqnarray*}
\frac{d\bf{B}}{dt} &=& \nabla \times \left[ (\bf{v} \times \bf{B})
- \eta \nabla \times \bf{B} \right] \\
&=& \nabla \times  (\bf{v} \times \bf{B})
+
\nabla \eta \cdot ( \sqrt{\eta} \nabla \BB)
- \sqrt{\eta} \left[ \nabla \times \left(\sqrt{\eta} \,
\nabla \times \bf{B} \right) \right]\\
\end{eqnarray*}  
or
\begin{equation}
\label{eqn1new}
\frac{d\bf{B}}{dt}
= \nabla \times (\bf{v} \times \bf{B})
+ ( \bf{U_{\eta}} \cdot  \nabla' \BB)
-
\nabla' \times \nabla' \times \bf{B} .
\end{equation}
\begin{figure}[ht]
\centering
\includegraphics[width=12.0cm]{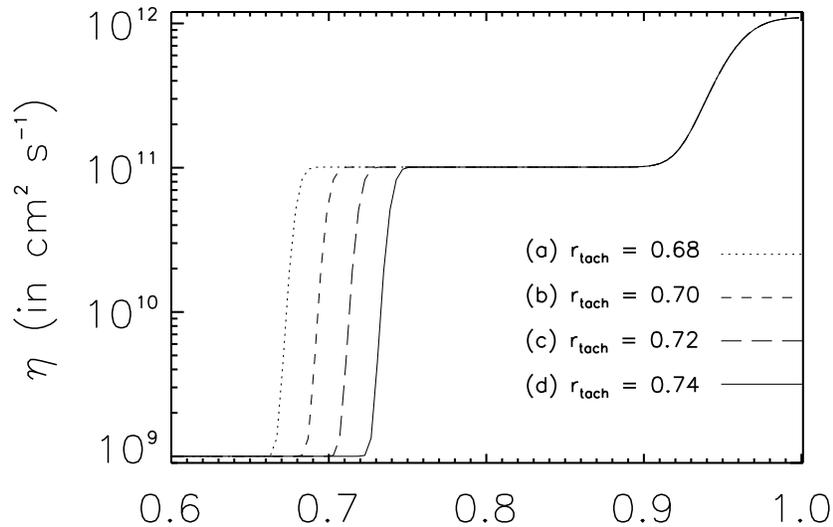}
\caption{Model diffusivity gradients in the solar convection zone. Parallel radial profiles used in solar dynamo simulations.  Tachocline is 0.7 r/R, photosphere at 1.0 (r=R).
Gradients are radially outward (to the right).
Curve (a), dotted line: steep diffusivity gradient above the tachocline. 
Curve (d), solid line:  steep diffusivity gradient below the tachocline. 
}
\end{figure}

\section{Physical considerations}
It is well known that magnetic fields diffuse more readily into
(or out of) regions of higher magnetic diffusivity (depending on relative magnetic
field concentrations), and that they
tend to ``freeze'' into regions of low diffusivity (high conductivity).
The new ``flow'' term in our induction/diffusion equation,
$( \bf{U_{\eta}} \cdot  \nabla' \BB)$, takes the form of
a convective derivative term,
with field advection velocity $\bf{U_{\eta}} \equiv \nabla \sqrt{\eta}$.
The last term $ - \nabla' \times \nabla' \times \bf{B}$
in (\ref{eqn1new}) is analogous to that in a heat flow equation; here
diffusion depends on the $\eta$-scaled gradient, $\nabla' = \sqrt{\eta} \, \nabla \,$
(Fig.1).
In nonlinear dynamic systems, the relationship may not be so simple, as the
diffusivity itself can vary in response to field changes.
Whether fields are advected into or out of regions of changing magnetic
diffusivity (or field strength) also depends on what other effects are operating
(e.g. magnetic buoyancy, magnetic reconnection, etc.), 
and with what relative strengths.

For example, we have observed in flux-transport dynamo simulations that, when magnetic fields
encounter steep diffusivity gradients leading into
regions of very low diffusivity,
e.g. near the tachocline,
it is possible for these fields to get trapped in high-conductivity regions, 
where they are prevented from further participation in meridional circulation (Fig.2).
In some cases, this trapping can quench the solar
dynamo altogether (Zita 2010).

Diffusion gradients due to magnetic reconnection have been shown to modify the
effective convection of the magnetic field, in simulations of the photospheric meridional flow \citep{cohen06}.
It has been shown \citep{BS} that gradients in resistivity (or diffusivity) can provide a
driving force for resistive tearing modes in regions of antiparallel magnetic fields.
This has been demonstrated in laboratory plasmas such as the
Reversed-Field Pinch \citep{Sarff}, and inferred in solar and galactic dynamics
(Parker 1970, 1971).  Magnetic quenching of turbulence has also been studied in supernova remnants (Ziegler 1996).
\begin{figure}[ht]
\centering
\mbox{(a)
     \includegraphics[height=5.5cm]{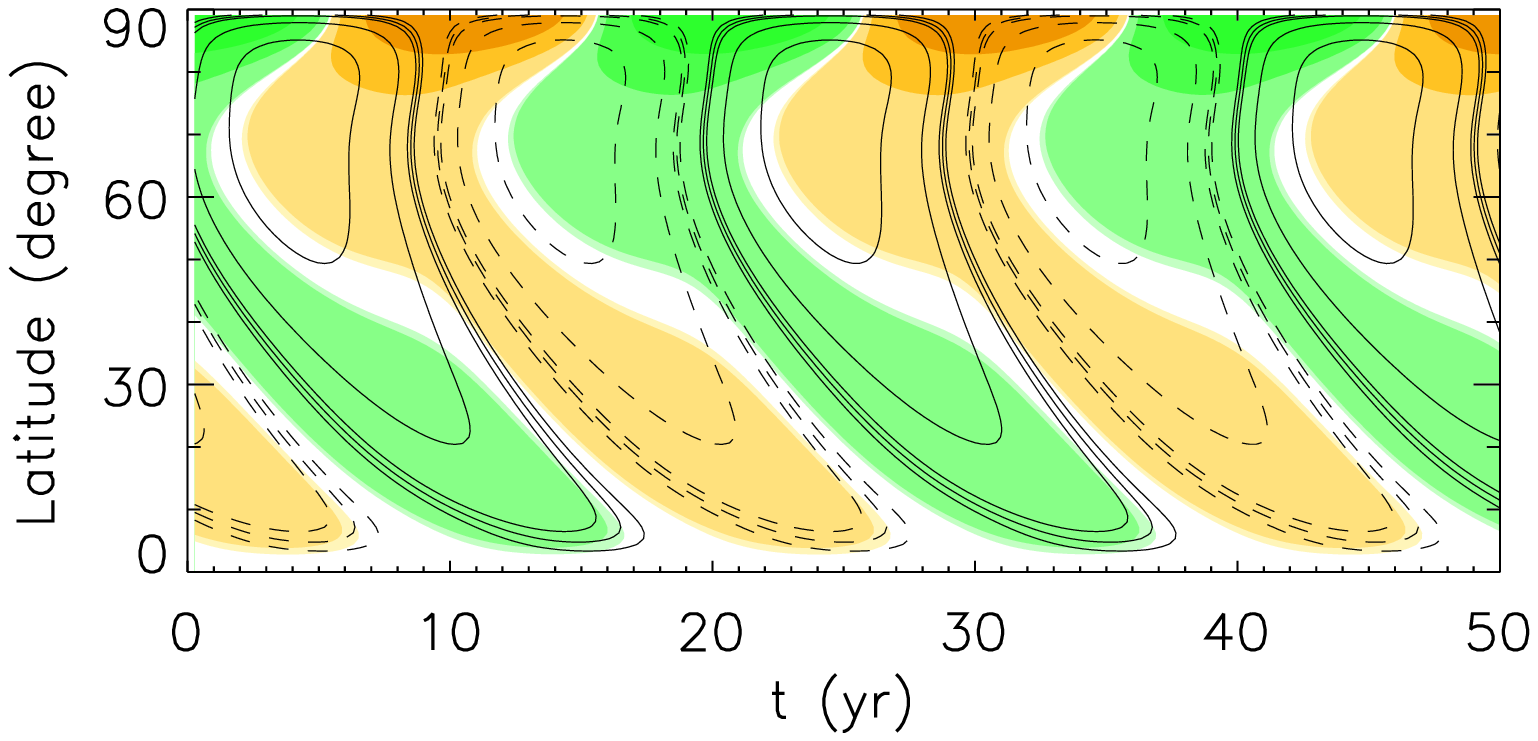}
     \includegraphics[height=4.0cm]{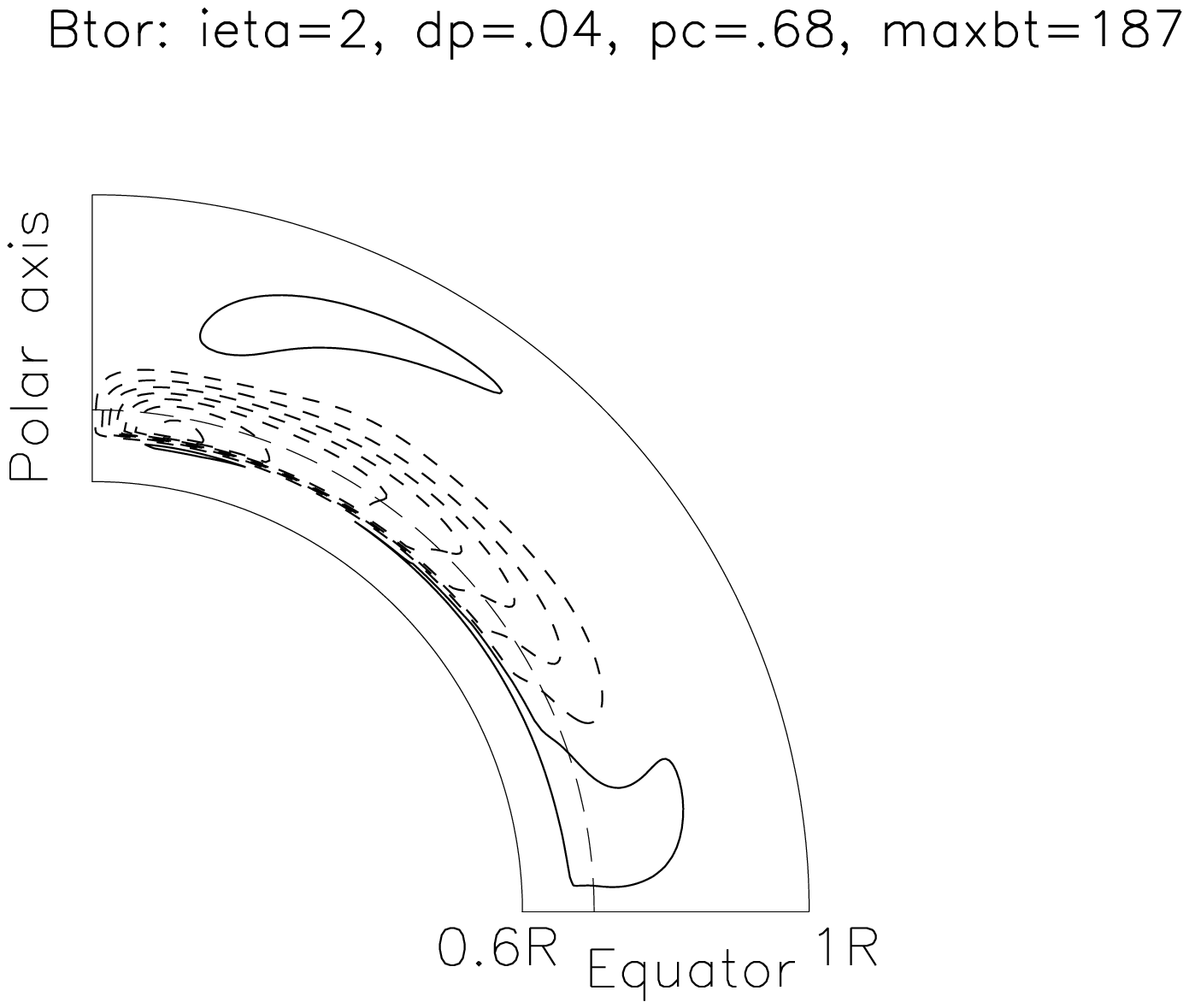}
     }
\mbox{(b)
    \includegraphics[height=5.5cm]{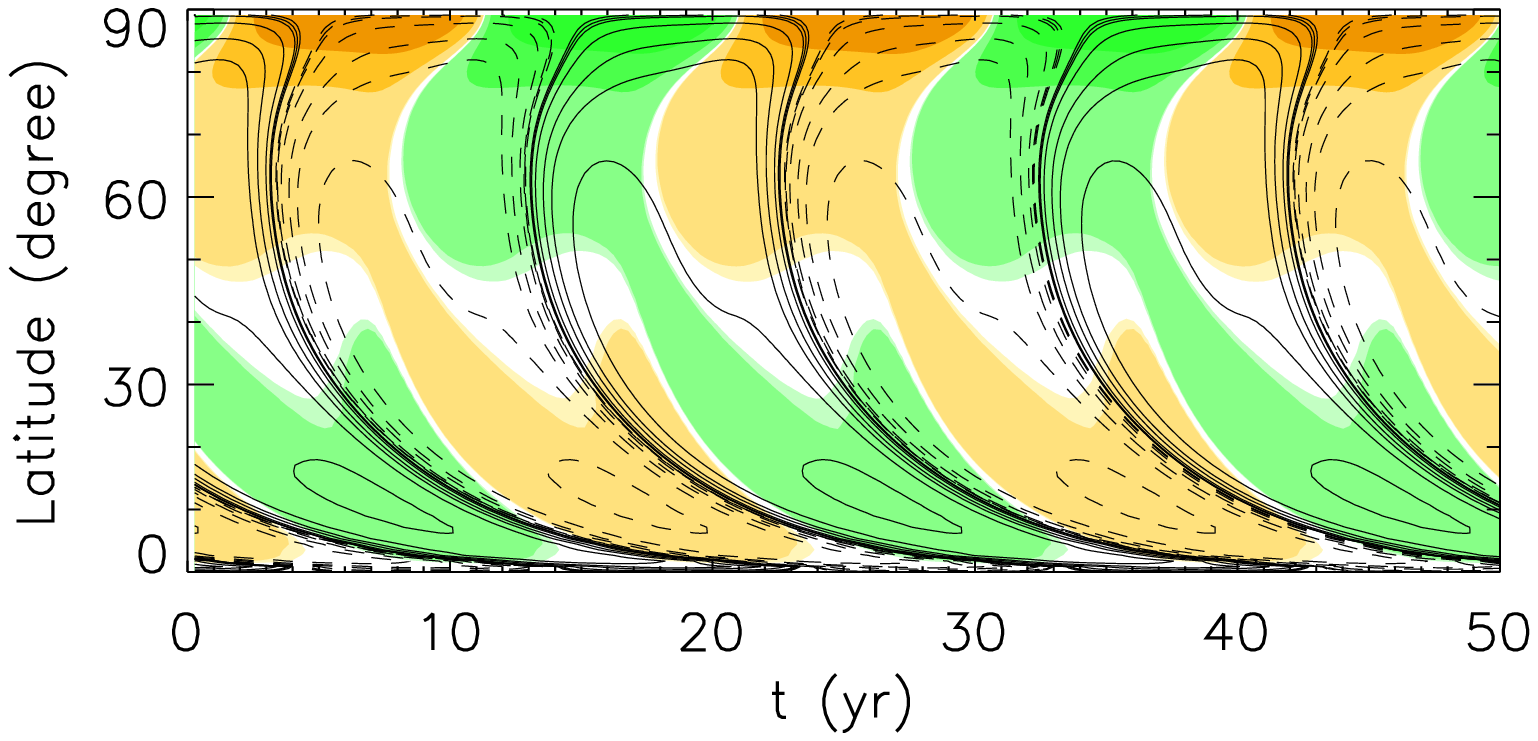}
     \includegraphics[height=4.0cm]{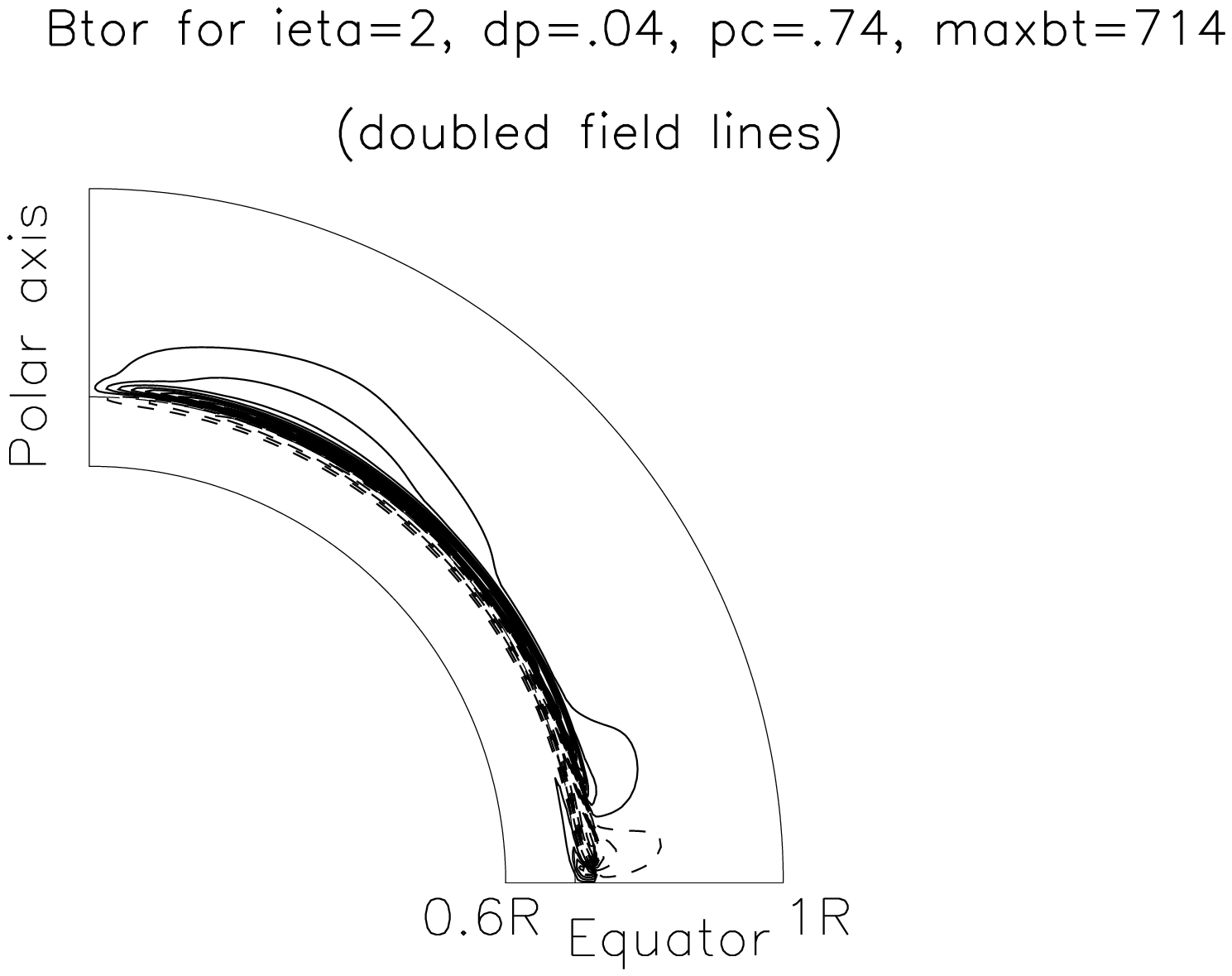}
     }
\caption{Results of dynamo simulation using profile in Fig.1a, 
a steep diffusivity gradient above the tachocline. Normal butterfly diagram (left) 
and normal toroidal flux patterns in convection zone (right) indicate normal solar dynamo.  
(b) Results of simulation using profile in Fig.1d, a steep diffusivity gradient below the tachocline. 
Pathological butterfly diagram (left) and trapped toroidal flux near the tachocline (right) 
indicate abnormal dynamo.}
\end{figure}

\section{Comparison to Mean Field Theory}
In equation (\ref{rudMF}), as usual in mean field theory (MFT), one considers 
two components of the total magnetic field ($\BB = \Bbar + \bp$).
The mean field $\Bbar$ varies on length
scales much longer than those of the field fluctuations $\bp$.
Similarly, the total velocity $\bf{v} = \vbar + \vp$ has
mean ($\vbar$) and fluctuating ($\vp$) components.
The expansion of the general induction equation (\ref{eqn1}) in mean field theory
\begin{eqnarray*}
\frac{d}{dt}(\Bbar+\bp) = \nabla \times \left[
(\vbar + \vp) \times (\Bbar + \bp)
- \eta \times \nabla \times (\Bbar + \bp) \right]
\end{eqnarray*}
is generally simplied to
\begin{equation}
\label{mainMF}
\frac{d\Bbar}{dt}=\nabla \times \left[ \vbar \times \Bbar
- \eta \times \nabla \times \Bbar
+ \curle \right] .
\end{equation}
One general result of mean field theory is that the $<\mathrm{volume \, \, \, average}>$ of the cross product between
small fluctuations in flows and small fluctuations in fields can yield a mean
electromotive force,
\begin{equation}
\label{emf}
\curle = \, <\vp \times \bp> =
\, \alpha \cdot \Bbar
+ \bf{\gamma} \times \Bbar,
\end{equation}
even if the local volume average of each fluctuating term vanishes.  These can produce significant emfs $\curle$, which can create and/or sustain mean
magnetic fields.  For example, in toroidal Reversed Field Pinch (RFP) plasma experiments \citep{Sarff}, 
the fluctuations
$<\mathrm{v_r b_z - v_z b_r}>$
create an $\curle_{\theta}$ which supports the mean axial $\BB_z$ well beyond the
predicted lifetimes of these high magnetic energy density devices, 
which operate outside the Kruskal-Shafranov limit 
and tend to spontaneously relax to a dynamic force-free state \citep{taylor74, taylor86}.
The $\alpha$ and $\bf{\gamma}$ terms in (\ref{emf}) are an alternate way of looking at the mean-field emf, 
using the diagonal and off-diagonal terms of the
turbulent velocity tensor, respectively \citep{KR}.
To complicate matters, Ziegler (1996) notes that each component of these tensors
should have its own quenching factor (Ferriere \& Schmidt 2000, p.139).

Returning to our original inquiry:  can our derivation help us understand the 
analytic origin of equation (\ref{rudMF})?
The diamagnetic pumping velocity in that relation, 
which ``will dominate the transport of mean magnetic fields'' (p.132),
is proportional to diffusivity gradients in the limiting cases of both weak and strong fields (p.114):
$U^{dia} = -\frac{1}{2} \nabla \eta_T$ when $\Bbar$ = 0, and
$U^{dia} = - \Phi^{dia}(\Omega,B) \gradeta$ for strong $\Bbar$, where
$\Phi^{dia}(\Omega,B) > 0$ is a function of rotation and field strength.
This differs in general from our advection velocity $\bf{U_{\eta}} \equiv \nabla \sqrt{\eta}$.

Though \Rud's MFT equation (2)
\begin{eqnarray*}
\frac{d\Bbar}{dt} = \nabla \times \left[ \vbar \times \Bbar
+ \alpha \cdot \Bbar
-\sqet \, \nabla \times \left( \sqet \Bbar \right) \right]
\end{eqnarray*}
appears similar to our induction equation
\begin{eqnarray*}
\frac{d\bf{B}}{dt} &=& \nabla \times  (\bf{v} \times \bf{B})
+
\nabla \eta \cdot ( \sqrt{\eta} \nabla \BB)
- \sqrt{\eta} \left[ \nabla \times \left(\sqrt{\eta} \,
\nabla \times \bf{B} \right) \right] \, ,\\
\end{eqnarray*}  
in fact these are quite independent.  
The former is from MFT and the latter is derived from MHD.
The advection velocity is field-dependent and linear with diffusivity gradients in MFT; 
while the advection velocity is field-independent and nonlinear with $\nabla \eta$ in MHD.  
Though we have not met our original goal, we have answered a potentially interesting new question.

\section{Summary and Future Work}

A magnetohydrodynamic analysis motivated by a question from mean field theory
has produced a new magnetic field diffusion equation (\ref{10de})
and a new form of the induction equation (\ref{eqn1new}).
These specify the role of magnetic diffusivity in changing magnetic fields.  
Magnetic field advection by gradients in turbulent diffusivity is relevant to solar dynamo theory, 
and may be fruitfully applied to space and laboratory plasmas.

\acknowledgments
\section{Acknowldegments}
Zita is grateful to Tom Bogdan for insightful discussions of the mathematical derivation
in Section~\ref{bozomath}, and to Mausumi Dikpati and Gustavo Guerrero for
a discussion about mean-field dynamo theory.

This work was supported by NSF grant 0807651, and partly by the Sponsored Research Program
of The Evergreen State College.

\clearpage

\end{document}